\documentclass[12pt]{article}
\usepackage{amsmath,amssymb,amsthm,amscd,graphicx}

\usepackage[all,cmtip]{xy}
\xyoption{line}

\input epsf.sty
\DeclareMathOperator{\Stokes}{\mathfrak{G}}

\renewcommand{\d}{\partial}
\newcommand{\WKB}{\mathbf{WKB}}
\newcommand{\med}{{\hbox{\tiny{med}}}}
\newcommand{\tun}{{\hbox{\tiny{tun}}}}

\newtheorem{theorem}{Theorem}[section]

\theoremstyle{definition}

\newtheorem{example}[theorem]{Example}

\newtheorem{remark}[theorem]{Remark}

\newcommand{\CF}{{\cal F}}
\newcommand{\CO}{{\cal O}}

\newcommand{\CE}{{\cal E}}

\newcommand{\tr}{{\rm Tr}}
\newcommand{\re}{{\rm e}}
\newcommand{\ri}{{\rm i}}
\newcommand{\rd}{{\rm d}}

\newcommand{\be}{\begin{equation}}
\newcommand{\ee}{\end{equation}}
\newcommand{\ba}{\begin{aligned}}
\newcommand{\ea}{\end{aligned}}
\newcommand{\ben}{\begin{eqnarray}\displaystyle}
\newcommand{\een}{\end{eqnarray}}

\newcommand{\sectiono}[1]{\section{#1}\setcounter{equation}{0}}

\newdimen\tableauside\tableauside=1.0ex
\newdimen\tableaurule\tableaurule=0.4pt
\newdimen\tableaustep
\def\phantomhrule#1{\hbox{\vbox to0pt{\hrule height\tableaurule width#1\vss}}}
\def\phantomvrule#1{\vbox{\hbox to0pt{\vrule width\tableaurule height#1\hss}}}
\def\sqr{\vbox{%
  \phantomhrule\tableaustep
  \hbox{\phantomvrule\tableaustep\kern\tableaustep\phantomvrule\tableaustep}%
  \hbox{\vbox{\phantomhrule\tableauside}\kern-\tableaurule}}}
\def\squares#1{\hbox{\count0=#1\noindent\loop\sqr
  \advance\count0 by-1 \ifnum\count0>0\repeat}}
\def\tableau#1{\vcenter{\offinterlineskip
  \tableaustep=\tableauside\advance\tableaustep by-\tableaurule
  \kern\normallineskip\hbox
    {\kern\normallineskip\vbox
      {\gettableau#1 0 }%
     \kern\normallineskip\kern\tableaurule}%
  \kern\normallineskip\kern\tableaurule}}
\def\gettableau#1{\ifnum#1=0\let\next=\null\else
\squares{#1}\let\next=\gettableau\fi\next}

\newcommand{\figref}[1]{Fig.~\protect\ref{#1}}

\begin{document}

\begin{titlepage}
\vskip .6cm

\centerline{\Large \bf
Multi-instantons in large $N$ Matrix Quantum Mechanics}

\medskip

\vspace*{4.0ex}

\centerline{\large \rm
Marcos Mari\~no$^{a,b}$ and Pavel Putrov$^a$}

\vspace*{4.0ex}

\centerline{ $^a$Section de Math\'ematiques and $^b$D\'epartement de Physique Th\'eorique}
\centerline{University of Geneva, Geneva, CH-1211 Switzerland}
\vspace*{2.0ex}
\centerline{marcos.marino@unige.ch}
\centerline{pavel.putrov@unige.ch}

\vspace*{15.0ex}
\centerline{\bf Abstract} \bigskip

We calculate the multi-instanton corrections to the ground state energy in large $N$ Matrix Quantum Mechanics. We find that they can be obtained, through a non-perturbative difference equation, from the multi-instanton series in conventional Quantum Mechanics, as determined by the exact WKB method. We test our results by verifying that the one-instanton correction  
controls the large order behavior of the $1/N$ expansion in the quartic potential and in the $c=1$ string.

\end{titlepage}

\vskip .6cm

\tableofcontents

\sectiono{Introduction}
The $1/N$ expansion of $U(N)$ gauge theories plays a central role in our current 
understanding of nonperturbative gauge theory dynamics. Most importantly, in some cases this expansion can be reinterpreted as 
a genus expansion in a dual string theory, leading to a connection between gauge theories and gravity theories. 

The $1/N$ expansion is in general an asymptotic expansion, and it has non-perturbative corrections of the form $\CO(\re^{-N})$ due 
to large $N$ multi-instantons. These corrections can trigger large $N$ phase transitions \cite{neuberger,gma} and, by general arguments \cite{lgzj}, they should control the 
large order behavior of the $1/N$ expansion. Moreover, when the gauge theory has a 
string dual, they can be reinterpreted in terms of D-branes. In spite of their relevance, 
the calculation of exponentially small corrections  to the $1/N$ expansion has not been pursued in detail, even in 
exactly solvable, low dimensional models. 

The simplest toy models for the $1/N$ expansion are matrix models and Matrix Quantum Mechanics. Both models were solved at the planar level in the pioneering 
paper by Br\'ezin, Itzkyson, Parisi and Zuber \cite{bipz}, and they have found an increasingly larger range of applications. 
Matrix models are now ubiquitous in physics, with important applications to 
two-dimensional gravity and topological string theory. However, explicit 
formulae for non-perturbative corrections in generic (off-criticality) matrix models have only appeared quite recently \cite{kk,mswone,mmnp,mswtwo,ps}. 

Matrix Quantum Mechanics has also been useful in many different areas. It provides for example 
a nonperturbative definition of $c=1$ strings \cite{gm} and of one-dimensional superstrings \cite{hat,toumbas}, and it has been instrumental in 
understanding some aspects of the AdS/CFT correspondence \cite{berenstein}. 
The purpose of this paper is to present explicit formulae for 
multi-instanton corrections in Matrix Quantum Mechanics, by focusing on the energy of the ground state, 
which is the basic observable of the theory.  
The first step in calculating these corrections is to find a useful characterization of the $1/N$ expansion to all orders. The next-to-leading correction to the planar 
result of \cite{bipz} was found in \cite{horder}, and it is not difficult to work out the full expansion. It turns out that, as in the case of matrix models \cite{mmnp}, the ground state energy 
is determined by a difference equation. This equation relates the Matrix Quantum Mechanics problem, at all orders in the $1/N$ expansion, to the WKB expansion of the ground state energy in standard Quantum Mechanics, and for the same potential. We can then use the beautiful results obtained in nonperturbative Quantum Mechanics, specially in \cite{zjone,ddp,zjj}, 
in order to extract the full large $N$ multi-instanton series, at all loops, in Matrix Quantum Mechanics. 

To illustrate our method, we analyze in detail Matrix Quantum Mechanics in a 
quartic potential, and we verify numerically that the one-instanton amplitude computed with our methods (at one loop) 
controls the large order behavior of the $1/N$ expansion. We also consider the double-scaling limit of this potential and reproduce in this way the large order behavior of the free energies 
of the $c=1$ string.  

The paper is organized as follows. In section 2 we review the exact WKB method in quantum mechanics, including multi-instanton corrections, following mostly 
the results of \cite{ddp}. In section 3 we start with a review of large $N$ Matrix Quantum Mechanics, and we reformulate the $1/N$ expansion in terms of a 
difference equation. We then proceed to the determination of the multi-instanton 
corrections, and finally we analyze the connection to large order behavior. We conclude with some open problems. In the Appendix we include for completeness a short review of the 
nonperturbative treatment of the Schr\"odinger equation of \cite{ddp}. 

\sectiono{Exact quantization conditions in Quantum Mechanics}

As we will see in the next section, the $1/N$ expansion in matrix quantum mechanics, as well as its non-perturbative corrections, can be calculated in a rather simple way by using 
the nonperturbative WKB method developed in \cite{zjone, voros, zjj,ddp,ddptwo}. In this section we review some relevant results from this method. 
 
 \subsection{Perturbative quantization conditions}
Let us consider the standard time-independent Schr\"odinger equation
\be
\hbar^2 \varphi''(x) +p^2(x,E) \varphi(x)=0, \qquad p(x,E) ={\sqrt{2(E-V(x))}}. 
\ee
If we write the wavefunction as
\be
\label{yansatz}
\varphi(x)=\exp \biggl[ {\ri \over \hbar} \int^x Y(x') \rd x' \biggr]
\ee
we transform the Schr\"odinger equation into a Riccati equation
\be
 Y^2(x)-\ri \hbar\frac{\rd Y(x)}{\rd x}=p^2(x,E), 
\end{equation} 
which we solve in power series in $\hbar$:
\be
Y(x, E, \hbar)=\sum_{k=0}^\infty Y_k(x,E)\hbar^k. 
\ee
The functions $Y_k(x,E)$ can be computed recursively as 
\begin{equation}\ba
 Y_0(x,E)&=p(x,E), \\
  Y_{n+1}(x,E)&=\frac{1}{2Y_0(x,E)}\left(\ri\frac{\rd Y_n(x,E)}{\rd x}-\sum_{k=1}^n Y_k(x,E)Y_{n+1-k}(x,E)\right).
  \ea
 \ee
 If we split $Y(x, E,\hbar)$ into even and odd powers of $\hbar$, 
 \be
 Y(x, E,\hbar) = Y_{\rm odd} (x, E,\hbar) + P(x,E, \hbar^2),
\end{equation} 
we find that 
\be
Y_{\rm odd}(x, E,\hbar)={\ri \hbar \over 2} {P'(x, E,\hbar^2) \over P(x, E,\hbar^2)}, 
\ee
and the wavefunction reads
\be
\varphi(x, E,\hbar)={1\over {\sqrt{P(x, E,\hbar^2)}}} \re^{{\ri \over \hbar} \int^x P(x', E,\hbar^2) \rd x' }. \label{WKB_wave}
\ee
\begin{figure}[!ht]
\leavevmode
\begin{center}
\includegraphics[height=4cm]{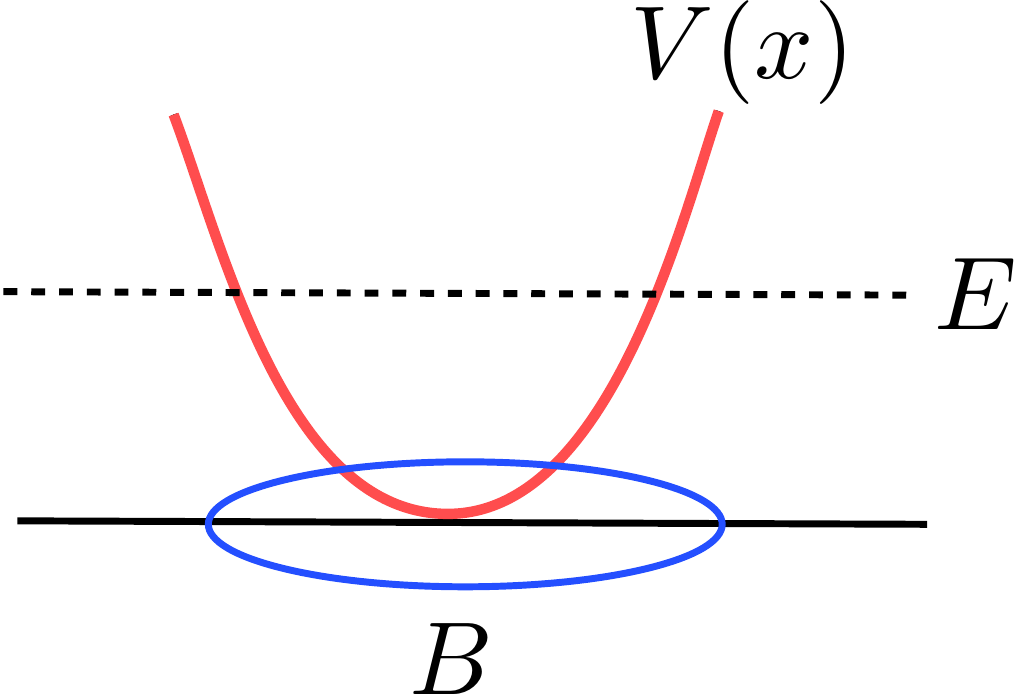}
\end{center}
\caption{A contour encircling the classical turning points.}
\label{Bcycle}
\end{figure} 

Let $x_*$ be a relative minimum of the potential $V(x)$, and let us consider an energy $E$ such that there are two roots of $E-V(x)=0$ near $x_*$. These are classical turning 
points for the potential. Let $B$ be a contour that encircles these roots, as depicted in \figref{Bcycle}. The classical frequency for the oscillation between these two turning points is
\be
\omega_B(E)= \oint_B p(x,E) \rd x
\ee
and the Bohr--Sommerfeld quantization condition reads
\be
\label{bs}
\omega_B(E_n)= 2\pi \hbar \Bigl(n-{1\over 2}\Bigr), \qquad n=1, 2, \cdots. 
\ee
It will be convenient in the following to write 
\be
\label{chiq}
\chi=\hbar \Bigl(n-{1\over 2}\Bigr)
\ee
so that the solution to (\ref{bs}) is given by a function $E_0^{(0)}(\chi)$. 

\begin{example} For the quartic potential
\be
\label{gquartic}
V(\lambda) = {1\over 2}\lambda^2 +g \lambda^4 
\ee
with $g>0$, the Bohr--Sommerfeld frequency is given by 
\be
\label{omegaquartic}
\omega_B(E,g)= \int_{-a}^{a} \rd 
\lambda{\sqrt {2E -\lambda^2 - 2g \lambda^4}},
\ee
where $a$, $-a$ are the classical turning points given by
\be
a^2=\frac{\sqrt{16 E g+1}-1}{4 g}.
\ee
The integral in (\ref{omegaquartic}) can be explicitly computed in terms of the elliptic functions $E(k)$, $K(k)$ as
\be
\omega(E,g)={4\over 3} (2g)^{1\over 2} (a^2+b^2)^{1\over 2}  \Bigl[ b^2 K(k) + (a^2-b^2) E(k)\Bigr]
\ee
where
\be
b^2=\frac{\sqrt{16 E g+1}+1}{4 g}, \qquad k^2={a^2 \over a^2 + b^2}.
\ee
The Bohr--Sommerfeld quantization condition leads to the energy levels
\be
\label{ekWKB}
E(\chi) =\chi + {3 \over 2} g\chi^2-{17\over 4} g^2\chi^3 +\CO(g^4)
\ee
as a power series in $g$. 
\end{example}

The condition (\ref{bs}) only incorporates the leading order WKB solution. An all-orders quantization condition was first proposed by Dunham in \cite{dunham}, and it reads as follows. 
We first define
\be
 \Omega_{\rm p}(E,\hbar)=\oint_B P(x, E, \hbar^2) \rd x= \omega_B(E)+\sum_{n=1}^\infty \omega^{(n)}_B(E)\hbar^{2n}, 
 \label{QC_pert}
 \ee
 where
 \be
  \omega^{(n)}_B(E)=\oint_B \rd x \, Y_{2n}(x,E).
  \ee
The all-orders quantization condition reads
\be
\label{allorders}
 \Omega_{\rm p}(E^{(0)}(\chi, \hbar),\hbar)=2\pi \chi, 
 \ee
 and the solution is a power series in $\hbar$, 
 \be
 E^{(0)}(\chi, \hbar)=\sum_{n=0}^{\infty}E^{(0)}_n(\chi) \hbar^{2n}. 
 \ee
The discrete energy levels are obtained by setting $\chi$ to its quantized values (\ref{chiq}). 
  
  \begin{example} In some cases, the quantization condition (\ref{allorders}) makes possible to compute the exact energy levels \cite{bow}. Let us consider the 
potential
\be
\label{kdvpot}
V(x)=-{V_0 \over \cosh^2 x} 
\ee
In this case, 
\be
\omega_B(E)=2\pi \Bigl({\sqrt{2 V_0}} - {\sqrt{-2 E}}\Bigr), \qquad \omega^{(n)}_B(E)= 2 \pi {\sqrt{2 V_0}} { {1\over 2} \choose n} \Bigl( {1\over 8 V_0}\Bigr)^n, 
\ee
The infinite sum appearing in the l.h.s. of (\ref{allorders}) can be summed up by using that
\be
{\sqrt{2 V_0}}\sum_{n=0}^{\infty} { {1\over 2} \choose n} \Bigl( {\hbar^2\over 8 V_0}\Bigr)^n =\Bigl( 2 V_0 +{\hbar^2 \over 4} \Bigr)^{1\over 2}
\ee
and the quantization condition gives the exact energy levels 
\be
\label{exacte}
E(\chi, \hbar)=-{1\over 2} \biggl[ \chi - \Bigl( 2 V_0 +{\hbar^2 \over 4} \Bigr)^{1\over 2}\biggr]^2. 
\ee
\end{example}

\subsection{Nonperturbative quantization conditions}

Nonperturbative corrections to energy levels in Quantum Mechanics can be obtained from a nonperturbative version of the quantization condition (\ref{allorders}). 
This condition involves a function 
\be
\label{npO}
\Omega(E,\hbar)=\Omega_{\rm p}(E,\hbar)+\Omega_{\rm np}(E,\hbar) 
\ee
where the perturbative piece is given by (\ref{QC_pert}), and $\Omega_{\rm np}(E,\hbar)$ is nonperturbative in $\hbar$. The exact quantization condition reads
\be
\label{npQC}
 \Omega(E(\chi, \hbar),\hbar)=2\pi \chi. 
\end{equation} 
and it results in nonperturbative corrections to the energy levels. Physically, $\Omega_{\rm np}(E,\hbar)$ is due to multi-instantons. 

The calculation of $\Omega_{\rm np}(E,\hbar)$ has been developed in a systematic and rigorous way in the context of the theory of resurgence \cite{ddp}, building on previous work 
(notably by J. Zinn--Justin) on multi-instantons in Quantum Mechanics \cite{zjone,zjj}. In the Appendix we summarize the approach of \cite{ddp}, which is quite elaborated. In this section we will 
illustrate the general method by considering standard situations where multi-instantons play a r\^ole: the case of an unstable potential (or false vacuum), and the case of a double-well potential. 

In general, $\Omega(E,\hbar)$ can be expressed in terms of the so-called {\it Voros multipliers} 
$a^{\gamma}$. A Voros multiplier is labelled by a contour $\gamma$ on the Riemann surface of the multivalued function of $x$, $p(x,E)$. It is defined by
\begin{equation}
 a^{\gamma}=\exp\Bigl\{ \frac{\ri}{\hbar}\oint_\gamma P(x,E,\hbar^2) \rd x\Bigr\}.
\end{equation} 
and we write
\begin{equation}
 \oint_\gamma P(x,E,\hbar^2)=\omega_{\gamma} +\sum_{n=1}^{\infty} \omega^{(n)}_{\gamma}(E)\hbar^{2n}, \qquad \omega_\gamma^{(n)}(E)=\oint_\gamma Y_{2n}(x,E) \rd x \;.
\end{equation}

\begin{figure}[!ht]
\leavevmode
\begin{center}
\includegraphics[height=4cm]{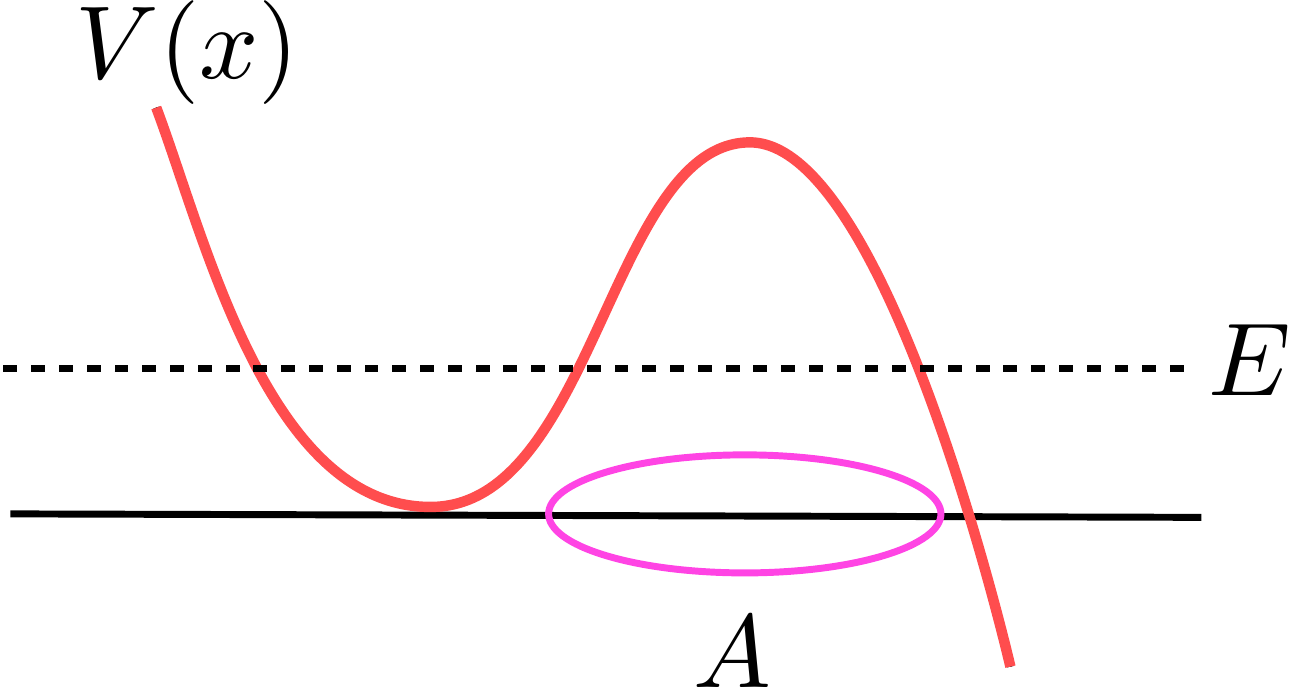}
\end{center}
\caption{The $A$ cycle goes from the turning point 
at the unstable minimum to the turning point of the instability.}
\label{unstableVoros}
\end{figure}

Let us first consider a false quantum-mechanical vacuum, like the one depicted in \figref{unstableVoros}. In this kind of situation, as it is well-known, 
the energy levels develop an imaginary part which reflects the instability. This imaginary part is nonperturbative in $\hbar$, and in order to compute it we have 
to compute the full $\Omega(E,\hbar)$ in (\ref{npO}). 
The perturbative part (the same as (\ref{QC_pert})) can be written in terms of Voros multipliers as
\be
\label{puns}
 \Omega_{\rm p}(E,\hbar)=\frac{\hbar}{\ri} \log{a^B}.
 \ee
Let us assume that the potential has an instability associated to an extra turning point, as depicted in \figref{unstableVoros}. Let us call $A$ 
the cycle in the Riemann surface going from the turning point 
at the unstable minimum to the turning point of the instability. The nonperturbative correction to $\Omega(E, \hbar)$ can be obtained by using the techniques developed in 
\cite{ddp} and summarized in the Appendix. It can be written in terms of the Voros multiplier for the $A$-cycle as \cite{ddp}
\be
\label{omnp}
\Omega_{\rm np}(E, \hbar) =\frac{\hbar}{\ri}\log(1+a^A)=-\ri\hbar a^A\left(1-\frac{a^A}{2}+\frac{(a^A)^2}{3}+\ldots\right). 
\ee
At leading order, 
\be
a^A\approx \re^{-\omega_{A}/\hbar}, \qquad \omega_A=\oint_A \ri p(x) \rd x. 
\ee
Notice that $\omega_A$ is real and positive for real $E$. 

\begin{remark} As reviewed in the Appendix, the quantities $\Omega(E, \hbar)$ are defined by Borel resummations. In the case of an unstable potential, and it is well-known, the Borel 
transform has a singularity in the positive real axis and one has to use a prescription to avoid it. In the formulae above we use right Borel resummation as described in Appendix 
\ref{resurgence}. Thus $\Omega$ in (\ref{npQC}) is actually $\Omega^+$, corresponding to the right Jost symbol $J^+(E)=1+a^B(1+a^A)$. The left Jost symbol is $J^-(E)=1+a^B$ and is purely perturbative.
\end{remark}

The exact, nonperturbative quantization condition is then given by (\ref{npQC}), where the perturbative and the nonperturbative part are given, respectively, by (\ref{puns}) and (\ref{omnp}). 
Let us represent the solution of this condition as
\begin{equation}
 E(\chi,\hbar)=\sum_{\ell \ge 0}E^{(\ell)}(\chi,\hbar), 
 \label{E_correction}
\end{equation}
where 
\be
\label{enexpansions}
\ba
E^{(0)}(\chi, \hbar) &=E_{\rm p} (\chi, \hbar)=\sum_{n\ge 0} E^{(0)}_{n} (\chi) \hbar^{2n}, \\
E^{(\ell)}(\chi,\hbar)&= \hbar \re^{-\ell \omega_A(E^{(0)})/\hbar}\sum_{n\ge 0} E^{(\ell)}_{n}(\chi) \hbar^n. 
\ea
\ee
In this series, $E^{(\ell)}_{n}(\chi)$ is the $(n+1)$-th loop correction in the $\ell$-th instanton sector. $E^{(0)}_{0}(\chi)$ is 
for example the result one obtains from the Bohr--Sommerfeld quantization condition. 
The full solution for the energy levels is then a double series in $\hbar$, $\re^{-\omega_A/\hbar}$, which is an example of {\it trans-series} solution.

Let us calculate for example $E^{(1)}(\chi, \hbar)$ at leading order in $\hbar$. 
After inserting (\ref{E_correction}) into the equation (\ref{npQC}) and identifying the powers of $\exp(-\omega_A/\hbar)$, we obtain
\begin{equation}
 \frac{\rd\omega_B}{\rd E}(E^{(0)}_{0}(\chi)) E^{(1)}(\chi,\hbar)+\Omega_{\rm np}(E^{(0)}_{0}(\chi))+\cdots=0,
\end{equation}  
so that at leading order in $\hbar$ we find
\begin{equation}
\ba
 E^{(1)}(\chi,\hbar)&=\ri \hbar  \Bigl( {\rd\omega_B \over \rd E}\Bigr)^{-1}_{E=E^{(0)}_{0}(\chi)}\re^{-\frac{1}{\hbar}\omega_{A}(E^{(0)}_{0}(\chi))}\left(1+\mathcal{O}(\hbar)\right)\\ 
 &=\frac{\ri \hbar}{T(E^{(0)}_{0}(\chi))} \re^{-\frac{1}{\hbar}\omega_{A}(E^{(0)}_{0}(\chi))}\left(1+\mathcal{O}(\hbar)\right)
 \ea
 \label{firstcorr}
\end{equation} 
where 
\be
\label{classicalperiod}
T(E)={\rd \omega_B \over \rd E}=\oint_B {\rd x \over p(x,E)}
\ee
is the period of the classical trajectory of energy $E$ between the turning points. 

Higher order corrections to (\ref{firstcorr}) can be obtained recursively. In fact, it is possible to write down a 
compact and relatively explicit expression for the solution of (\ref{npQC}). The function $E(\chi)$ solves the equation 
\be
\omega_B(E)+\psi(E)=2\pi\chi
\ee
where
\be
\psi(E)=\omega^{\rm q}_B(E)+ \Omega_{\rm np}(E), \qquad \omega^{\rm q}_B(E)= \sum_{n=1}^\infty\hbar^{2n}\omega^{(n)}_B.
\ee
Then one can write
\be
\ba
 \frac{1}{2\pi}\,\frac{\d E(\chi)}{\d \chi}&=\frac{1}{\omega_B'(E)+\psi'(E)}=\oint\frac{\rd\lambda}{\omega_B(\lambda)
 +\psi(\lambda)-2\pi\chi}=\oint \re^{-\frac{\psi(\lambda)}{2\pi}\frac{\d}{\d\chi}}\,\frac{d\lambda}{\omega_B(\lambda)-2\pi\chi}\\
&=\,:\re^{-\frac{\d}{\d\chi}\frac{\psi(E^{(0)}_0(\chi))}{2\pi}}:\,\frac{1}{\omega_B'(E^{(0)}_0(\chi))}=\,:\re^{-{D}\psi(E^{(0)}_0)}:\,\frac{1}{T(E^{(0)}_0)}\\
&=\sum_{k=0}^\infty \frac{(-1)^k}{k!}\,D^k\,\left(\frac{\psi^k(E^{(0)}_0)}{T(E^{(0)}_0)}\right)
\label{Explicit_proof1}
\ea
\ee
where on the second line $E^{(0)}_0$ is the solution to the equation $\omega_B(E)=2\pi\chi$ (i.e. is the energy defined by the 
Bohr--Sommerfeld quantization condition), and 
\be
{D}=\frac{1}{2\pi}\frac{\d}{\d\chi}=\frac{1}{T(E^{(0)}_0)}\frac{\d}{\d E^{(0)}_0}.
\ee
Here the normal ordering $:\,:$ is defined by the rule ``all $D$'s to the left''. One can now act by ${D}^{-1}$ on both sides of (\ref{Explicit_proof1}) to obtain
 the explicit solution
\begin{equation}
 E(\chi)=\,:{D}^{-1}\,\re^{-{D}\psi(E^{(0)}_0)}:\,\frac{1}{T(E^{(0)}_0)}=\,:{D}^{-1}\,\re^{-{D}\omega^{\rm q}_B(E^{(0)}_0)}(1+a^A)^{\ri \hbar D}:\,\frac{1}{T(E^{(0)}_0)} \label{Explicit}
\end{equation} 
One can extract from this solution the expression for the $\ell$-instanton correction:
\begin{equation}
 E^{(\ell)}=\frac{\ri \hbar}{\ell!}\,P^{(\ell)}(\hbar D):\re^{-D\omega^{\rm q}_B}:\frac{(a^A)^\ell}{T}
\end{equation} 
where 
\begin{equation}
 P^{(\ell)}(x)=\left\{\begin{array}{cl}
                1\;,& \ell=1\;, \\ \prod\limits_{k=1}^{\ell-1}(\ri x-k)\;, & \ell>1\;.
               \end{array} \right.
\end{equation}
Since $\omega^{\rm q}_B=\mathcal{O}(\hbar^2)$, and
\be
(\hbar D)\re^{-\frac{\omega_A}{\hbar}}=\frac{T_\tun}{T}\re^{-\frac{\omega_A}{\hbar}}(1+\mathcal{O}(\hbar)),
\ee
where
\be
\;T_\tun(E)=-\frac{\d\omega_A(E)}{\d E}
\ee
we easily deduce the one-loop expression for a general multi-instanton correction $\ell\ge 1$:
\begin{equation}
  E^{(\ell)}_0=\frac{\ri}{T(E^{(0)}_0)\,\ell!}\,P^{(\ell)}\left(\frac{T_\tun(E^{(0)}_0)}{T(E^{(0)}_0)}\right)\;.
\end{equation}

\begin{figure}[!ht]
\leavevmode
\begin{center}
\includegraphics[height=3cm]{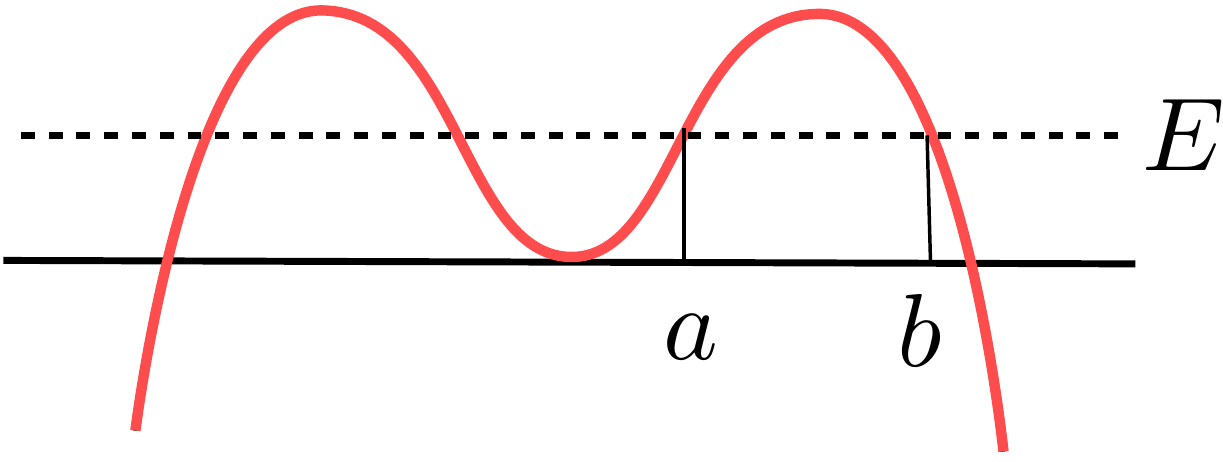}
\end{center}
\caption{The inverted quartic potential. The $A$ cycle goes from $a$ to $b$.}
\label{inver}
\end{figure} 
\begin{example}  Let us consider the inverted quartic potential (\ref{gquartic}) with $g=-\kappa<0$. In this case, $\lambda=0$ is a false vacuum, see 
\figref{inver}. We first write
\be
2E -\lambda^2 + 2 \kappa \lambda^4 = 2\kappa (a^2-\lambda^2)(b^2 - \lambda^2), 
\ee
where 
\be
a^2=\frac{-\sqrt{1-16 E \kappa}+1}{4 \kappa}, \qquad b^2=\frac{\sqrt{1-16 E \kappa}+1}{4 \kappa}.
\ee
$a, -a$ are the classical turning points, while $b,-b$ are the turning points of the instability. The relevant period integrals are, 
\be
\label{periodints}
\ba
\omega_B&=2(2\kappa)^{1/2}\int_{-a}^a {\sqrt{(a^2-\lambda^2)(b^2 - \lambda^2)}} \rd \lambda,\\
\omega_A&=2(2\kappa)^{1/2}\int_a^b {\sqrt{(\lambda^2-a^2)(b^2 - \lambda^2)}} \rd \lambda.
\ea
\ee
The periods can be computed with elliptic functions. The elliptic modulus is 
\be
\label{inversemod}
k^2={a^2 \over b^2},
\ee
and we obtain
\be
\label{periodex}
\ba
\omega_B&={4\over 3} (2\kappa)^{1\over 2} b \Bigl[ (a^2-b^2) K(k)+(a^2+b^2) E(k)\Bigr],\\
\omega_A&={2\over 3} (2\kappa)^{1\over 2} b  \Bigl[ (a^2+b^2) E(k')-2 a^2 K(k')\Bigr],
\ea
\ee
where as usual $k'^2=1-k^2$. Of course, since the potential is symmetric, there is another contribution to $\Omega_{\rm np}$. 
It comes from the Voros multiplier associated to the cycle 
which goes from $-a$ to $-b$. Therefore
\begin{equation}
 \Omega(E,\hbar)=\frac{\hbar}{\ri} \log{a^B}+2\frac{\hbar}{\ri}\log(1+a^A)\;.
\end{equation} 
To check this result, obtained with the WKB approximation, we can compare it to the leading order result (in $\kappa$) for the one-instanton correction to the energy of the 
$n$-th level. This is presented in e.g. \cite{ddp,UniTreat} and reads (we set $\hbar=1$)
\begin{equation}
 E^{(1)}_{n}(\kappa)=\frac{\ri}{\sqrt{2\pi}(n-1)!}\,\left(\frac{4}{\kappa}\right)^{n-1/2}\,\re^{-\frac{1}{3\kappa}} \, \bigl(1+\mathcal{O}(\kappa)\bigr) 
\label{pertinst}
\end{equation} 
To compare this with the result using the WKB method presented above, we have to keep all orders in $\hbar$, but only the leading terms in $\kappa$. We find that 
\be
\omega_A={1\over 3\kappa}-\chi\,\log\frac{\kappa \chi}{4\re} +\CO(\kappa), \quad \ri \omega^{(n)}_A ={c_n \over \chi^{2n-1}} +\CO(\kappa), 
\ee
where
\be
 c_1=\frac{1}{24}\,,\,c_2={\frac {7}{2880}}\,,\,c_3={\frac {31}{
40320}}\,,\,c_4={\frac {127}{215040}}\,,\,\ldots
\end{equation} 
The Voros 
multiplier $a^A$ is then given, in this limit, by
\be
a^A(\chi,\kappa)=\exp\left\{-\frac{1}{3\kappa}-\chi\,\log\frac{\kappa \chi}{4\re}+\sum_{n=1}^{\infty}\frac{c_n}{\chi^{2n-1}}+\mathcal{O}(\kappa)\right\},
\end{equation} 
and one finds
\be
 E^{(1)}(\chi, \kappa)=\frac{\ri}{2\pi}\,\left(\frac{4}{\kappa}\right)^{\chi}\,\re^{-\chi\,\log\frac{\chi}{\re}+\sum\limits_{n=1}^{\infty}\frac{c_n}{\chi^{2n-1}}}\, \re^{-\frac{1}{3\kappa}}\,(1+\mathcal{O}(\kappa))\;. \label{E_1inst_g_WKB}
\end{equation} 
One can check that
\begin{equation}
 -\log  \left( {\frac {\Gamma  \left( \chi+1/2 \right)}{
\sqrt {2\pi }}} \right) = -\chi\,\log\frac{\chi}{\re}+\sum\limits_{n=1}^{\infty}\frac{c_n}{\chi^{2n-1}}\;,\label{Stirling_m}
\end{equation}
as an asymptotic expansion in powers of $\chi$, 
and thus expressions (\ref{E_1inst_g_WKB}) and (\ref{pertinst}) coincide when we set $\chi=n-1/2$. 
\end{example}

\begin{figure}[!ht]
\leavevmode
\begin{center}
\includegraphics[height=4cm]{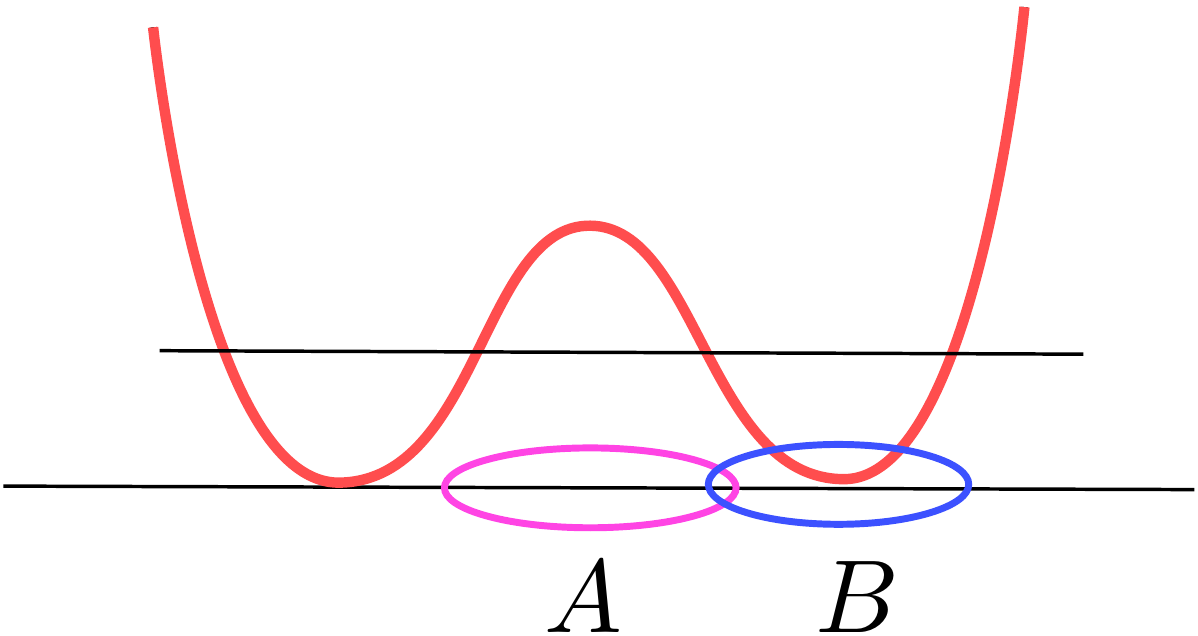}
\end{center}
\caption{The double well. The $A$ cycle goes from $-a$ to $a$.}
\label{doublewell}
\end{figure} 

It is possible to use the formalism of \cite{ddp} to study other cases, like the double-well potential shown in \figref{doublewell}. 
As it is well known (see for example \cite{zjj}), the energy levels split into even and odd levels, according 
to the symmetry properties of the wavefunctions, which we will denote by $E_{n,\pm}$, respectively.  Like before, the perturbative contribution to $\Omega$ is given by 
(\ref{QC_pert}), where the $B$ cycle encircles one of the degenerate minima of the potential. Define now
\be
Q= \frac{1+ \ri a^{A/2}}{\sqrt{1+a^A}}
\end{equation} 
where $A$ is the cycle shown in \figref{doublewell}. The exact energies $E_{\pm}(\chi, \hbar)$ are defined by the quantization conditions
\be
\Omega_\pm (E_{\pm}(\chi, \hbar))=2\pi \chi, 
\ee
where
\be
\label{omegapm}
\Omega_{\pm}(E)=\Omega_{\rm p}(E)\pm \ri \hbar \log Q. 
\ee
This is a compact way of encoding the non-perturbative quantization conditions for the double-well potential first conjectured by Zinn--Justin \cite{zjone}. One can 
check that the leading contribution to the energy difference coincides with the known answer
\begin{equation}
 E_+-E_-=\frac{2\hbar}{T}\,e^{-\frac{s(E)}{\hbar}}(1+\mathcal{O}(\hbar)),
\end{equation} 
where
\be
s(E)=\int_{-a}^a |p(x,E)|\rd x
\ee
and $-a,a$ are the turning points for the $A$ cycle. 
\begin{remark} In this case, the results are obtained by using the so-called median resummation described in Appendix \ref{resurgence}. The result of this resummation is manifestly real. The computation of the left and right Jost symbols can be found in \cite{ddp}. The median Jost symbol is 
\begin{equation}
 J^{\med}=2+(1+a^A)^\frac{1}{2}(a^B+a^{-B})\propto (1+\re^{\frac{i}{\hbar}\Omega_+})(1+\re^{\frac{i}{\hbar}\Omega_-})\;.
\end{equation} 
\end{remark}

\sectiono{Nonperturbative effects in large $N$ Matrix Quantum Mechanics}

\subsection{$1/N$ expansion of the ground state energy}

In Matrix Quantum Mechanics (MQM) the degrees of freedom are the entries of a Hermitian $N\times N$ matrix $M$, and the Euclidean Lagrangian is
\be
\label{LagM}
L_M  =\tr \Bigl[ {1\over 2}\dot M^2 - V_N(M) \Bigr], 
\ee
where $V_N(M)$ is a potential. Notice that this problem has a symmetry 
\be
M \rightarrow U M U^{\dagger}
\ee
where $U$ is a constant unitary matrix. MQM can be regarded as a one-dimensional field theory for a quantum field $M(t)$ taking values in the adjoint representation of $U(N)$.

As  first shown in \cite{bipz}, the ground state energy of MQM has a $1/N$ expansion which can be obtained in terms of a system of free fermions. In this 
section we will review the results of \cite{bipz} and we will extend them to all orders in the $1/N$ expansion. The Hamiltonian 
operator of MQM is given 
\be
H=\tr  \Bigl[ -{1\over 2}{\partial^2 \over \partial M^2} + V(M) \Bigr], 
\ee
where
\be
\tr {\partial^2 \over \partial M^2}=\sum_{ab}  
{\partial^2 \over \partial M_{ab} M_{ba}}
\ee
In order to study the spectrum of this Hamiltonian, it is useful to write the matrix $M$ as
\be
\label{Mdiag}
M=U \Lambda U^{\dagger}
\ee
where 
\be
\Lambda ={\rm diag}(\lambda_1, \lambda_2, \cdots, \lambda_N)
\ee
is a diagonal matrix. It is easy to show that (see for example \cite{affleck}) 
\be
\label{derlam}
\tr {\partial^2 \over \partial M^2}={1\over \Delta(\lambda)} \sum_{a=1}^N \Bigl( 
{\partial \over \partial \lambda_a}\Bigr)^2 \Delta(\lambda) + \sum_{a<b} {{\cal F}_{ab} \over 
(\lambda_a - \lambda_b)^2}, 
\ee
where 
\be
\Delta(\lambda)=\prod_{a<b} (\lambda_a -\lambda_b)
\ee
is the Vandermonde determinant, and 
$\CF_{ab}$ are differential operators w.r.t. the angular coordinates in $U$. 

Let us now consider {\it singlet} states. 
These are invariant under the $U(N)$ group, and in particular 
they depend only on the eigenvalues $\lambda_a$ up to permutation. The reason is that, after reduction to eigenvalues, 
the $U(N)$ group still acts through the Weyl group, i.e. by permuting eigenvalues. Therefore, singlet states 
will be represented by a symmetric function,
\be
\Psi(\lambda_i).
\ee
If we are now interested in computing 
the spectrum of the Hamiltonian for singlet states, we can reformulate the problem as a problem of $N$ fermions 
in the potential $V_N(\lambda)$. To see this, we introduce a completely antisymmetric wavefunction
\be
\Phi(\lambda) =\Delta(\lambda) \Psi(\lambda)
\ee
The equation 
\be
H\Psi =E  \Psi
\ee
can now be written as 
\be
\biggl( \sum_{i=1}^N h(\lambda_i) \biggr) \Phi(\lambda_j)=E  \Phi(\lambda_j)
\ee
where $h(\lambda)$ is the Hamiltonian
\be
h(\lambda)=-{1\over 2} {\partial^2 \over \partial \lambda^2} +V_N(\lambda).
\ee
Since the fermions are not interacting, we can just solve the Schr\"ondiger equation for a single particle of unit mass,
\be
h(\lambda) \phi_n(\lambda)= E_n \phi_n(\lambda).
\ee
In particular, the ground state of the system (in the singlet sector) will be obtained by putting the $N$ fermions in the first $N$ 
energy levels of the potential, and its energy will be
\be
\CE (N)=\sum_{n=1}^N E_n 
\ee

We want to compute the ground state energy $\CE(N)$ at large $N$, and as an expansion in $1/N$. 
In order to have a good large $N$ limit,  $V_N (\lambda)$ 
must be of the form \cite{horder}, 
\be
V_N(\lambda)=NV(\lambda/{\sqrt N}).
\ee
After rescaling
\be
\label{laresc}
\lambda \rightarrow N^{1\over 2} \lambda,
\ee
the Schr\"odinger problem becomes
\be
\label{hNeq}
\biggl\{ -{1\over 2N^2} {\rd^2 \over \rd \lambda^2} +V(\lambda) \biggr\}\phi_n(\lambda)=\hat e_n \phi_n(\lambda).
\ee
where we denoted
\be
\hat e_n={1\over N} E_n.
\ee
In this equation, $1/N$ plays the r\^ole of $\hbar$, and this suggests using the WKB approximation in the calculation of the energy levels 
$\hat e_n$. The total energy of the ground state is now
\be
\CE(N)=\sum_{k=1}^N E_k = N \sum_{k=1}^N \hat e_k. 
\ee

It will be convenient to re-scale the potential in such a way that the Schr\"odinger problem (\ref{hNeq}) becomes
\be
\label{hNeq2}
\biggl\{ -{ t^2 \over 2N^2} {\rd^2 \over \rd \lambda^2} +v(\lambda) \biggr\}\phi_n(\lambda)= t^{2-\alpha} \hat e_n \phi_n(\lambda).
\ee
The value of $\alpha$ depends on the potential, and $t$ will be later identified with the 't Hooft parameter. 

\begin{example} For the potential 
\be
\label{tquartic}
V(\lambda)={\lambda^2 \over 2} + t \lambda^4, 
\ee
we find the form (\ref{hNeq2}) after setting $\lambda \rightarrow \lambda/t$, and $\alpha=1$, and 
\be
v(\lambda)= {\lambda^2 \over 2} + \lambda^4.
\ee
For the potential (\ref{kdvpot}) with $V_0=1/t^2$, we have $\alpha=0$. 
\end{example}

We now set 
\be
 e_k =t^{2-\alpha} \hat e_k
\ee
and
\be
t=g_s N.
\ee
We will write the $1/N$ expansion of the ground state energy as
\be
\CE(t, g_s) =\sum_{g=0}^{\infty} g_s^{2g-2} \CE_g(t). 
\ee
The Schr\"odinger problem (\ref{hNeq}) becomes
\be
\label{hNeq3}
\biggl\{ -g_s^2 {\rd^2 \over \rd \lambda^2} +v(\lambda) \biggr\}\phi_n(\lambda)= e_n \phi_n(\lambda).
\ee
If we denote by 
\be
\chi={ n-{1\over 2} \over N}, \qquad z=g_s (n-1/2)= t \chi, 
\ee
we find that the all-orders perturbative WKB solution for the energy levels is given by
\be
\Omega_{\rm p}(e(z,g_s), g_s)= 2\pi z, 
\ee
and it defines the perturbative function 
\be
\label{epert}
e(z, g_s)=\sum_{n=0}^{\infty} e_n(z) g_s^{2n}.
\ee
For $z=t$, i.e. $\chi=1$ or equivalently $n\sim N$, the function $e_0(t)$ is the WKB approximation to the Fermi energy of the fermionic system \cite{bipz}. 
For future use, we will denote it by $e_F(t)$. 
\begin{example} For the potential (\ref{kdvpot}) we have
\be
\label{ekdv}
e(z,g_s)=-{1\over 2} \biggl[ z - \sqrt{2 + g_s^2/4}\biggr]^2. 
\ee
\end{example}

We now derive a general formula for $\CE(t,g_s)$, which generalizes \cite{bipz,horder} to all orders in the $1/N$ expansion. 
We first notice the following analogue of the Euler--Maclaurin asymptotic formula, 
\be
\label{eum}
\ba
  N\sum_{n=1}^N\phi\left(\frac{n-1/2}{N}\right)&=N\left.\sum_{i=1}^N \re^{\frac{n-1/2}{N}\,\frac{\rd}{\rd\chi}}\phi(\chi)\right|_{\chi=0}
=N^2\left.\left(\re^{{\rd}/{\rd\chi}}-1\right)\,\frac{\frac{1}{2N}}{\sinh\frac{\rd/\rd\chi}{2N}}\phi(\chi)\right|_{\chi=0} \\ 
&=
N^2\int\limits_0^1\phi(\chi)\rd\chi-\sum_{k=1}^\infty\, N^{2-2k}\,\frac{\left(1-2^{1-2k}\right)B_{2k}}{(2k)!}\left.\frac{\rd^{2k-1}\phi(\chi)}{\rd\chi^{2k-1}}\right|^{\chi=1}_{\chi=0}\;.
\ea
\ee
Define first
\be
\CF(t,g_s) =t^{1-\alpha} \CE(t,g_s). 
\ee
Then, one has that
\be
\CF(t,g_s)={N\over t} \sum_{n=1}^N e_n
\ee
and using (\ref{eum}) one finds
\be
\label{fber}
g_s^2 \CF(t,g_s) = \int_0^t \rd z \, e(z, g_s) -\sum_{k=1}^\infty g_s^{2k} \frac{\left(1-2^{1-2k}\right)B_{2k}}{(2k)!}\left.\frac{\rd^{2k-1}e(z,g_s)}{\rd z^{2k-1}}\right|^{z=t}_{z=0}.
\ee
We can then write (\ref{fber}) as a difference equation
\be
\label{exactrel}
g_s\Bigl\{ \CF\Bigl(t + {g_s\over 2} \Bigr) -\CF\Bigl(t -{g_s\over 2} \Bigr)\Bigr\}=e(t, g_s).
\ee
We then see that the large $N$ expansion of the ground state energy in MQM can be obtained, through (\ref{exactrel}), from the WKB expansion of the energies in an ordinary 
Quantum Mechanics problem with potential $v(\lambda)$. Moreover, (\ref{exactrel}) can be used as well to compute non-perturbative corrections, as we will see in a moment. 

In the case of a symmetric potential, like the double-well, 
the equation (\ref{exactrel}) has to be modified as follows:
\be
g_s\Bigl\{ \CF\Bigl(t + {g_s\over 2} \Bigr) -\CF\Bigl(t -{g_s\over 2} \Bigr)\Bigr\}=e_+(t, g_s)+e_-(t, g_s),
\ee
where $e_{\pm}(t,g_s)$ are obtained from the quantization condition (\ref{omegapm}). 

\begin{remark} The equation (\ref{exactrel}) is very similar, formally, to the equation determining the total free energy of the one-cut matrix model with the method of orthogonal polynomials, which 
can be also reformulated as a Toda-like difference equation \cite{mmnp}. The function 
$e(z,g_s)$ plays the r\^ole of the function $R(z, g_s)$, which is obtained as the continuum limit of the coefficients $r_n$ appearing in the recursion relation of 
orthogonal polynomials.
\end{remark}

\begin{example} For the potential $V(\lambda)$ in (\ref{kdvpot}) we find, by a direct calculation using the result (\ref{exacte}),
\be
\CE(t,g_s)= N\sum_{n=1}^N e_n= g_s^{-2}\biggl\{ -1-{t^2\over 6} + {t\over 2} {\sqrt{2+ {g_s^2\over 4}}} \biggr\} -{1\over 12}. 
\ee
In this case, $\alpha=0$ and
\be
\CF(t,g_s)=t \CE(t,g_s)
\ee
satisfies indeed (\ref{exactrel}) with $e(t,g_s)$ given by (\ref{ekdv}). For the quartic potential (\ref{tquartic}) we obtain from (\ref{ekWKB}) 
the power series expansion for the planar approximation, 
\be
\CE_0(t)={t^2 \over 2} +{t^3 \over 2} -{17t^4 \over 16}+ {75 t^5 \over 16} +\CO(t^6)
\ee
which is the classic result of \cite{bipz}.
\end{example}

\subsection{Nonperturbative corrections to the $1/N$ expansion}

The perturbative expansion of $\CE(t,g_s)$ (or, equivalently, to $\CF(t,g_s)$) can be obtained from (\ref{exactrel}) by plugging in the perturbative 
expansion of $e(t,g_s)$ (\ref{epert}). But the function $e(t,g_s)$ has non-perturbative corrections coming from the exact WKB quantization condition (\ref{npQC}). Therefore, we have
\be
\label{efull}
e(t,g_s)=\sum_{\ell\ge0} e^{(\ell)}(t,g_s), 
\ee
with
\be
\ba
e^{(0)}(t,g_s)&=\sum_{g=0}^{\infty} e_g(t) g_s^{2g}, \\
e^{(\ell)}(t,g_s)&=g_s \re^{-\ell \omega_A (e_F(t))/g_s} \sum_{n=0}^{\infty} e^{(\ell)}_n(t) g_s^{n},  \quad \ell\ge 1,
\ea
\ee
just as in (\ref{enexpansions}). 
In order to obtain the nonperturbative corrections to $\CF(t,g_s)$, we simply have to consider the difference equation (\ref{exactrel}) as an exact statement, and plug in the full expansion 
(\ref{efull}). This leads to a multi-instanton expansion for $\CF(t,g_s)$ of the form
\be
\CF(t,g_s)=\sum_{\ell\ge0}\CF^{(\ell)}(t,g_s)
\ee
where
\be
\label{cfexpansion}
\ba
\CF^{(0)}(t,g_s)&=\sum_{g=0}^{\infty} \CF_g(t) g_s^{2g-2}, \\
\CF^{(\ell)}(t,g_s)&= \re^{-\ell \omega_A (e_F(t))/g_s} \sum_{n=0}^{\infty} \CF^{(\ell)}_n(t) g_s^{n},  \quad \ell\ge 1.
\ea
\ee

To see how this works, let us calculate the one-loop, one-instanton correction $\CF^{(1)}_0(t)$ in the unstable potential of \figref{unstableVoros}. 
By plugging the expansion (\ref{cfexpansion}) in (\ref{exactrel}) we find, first of all, that 
the instanton action appearing in (\ref{cfexpansion}) is indeed $\omega_A(e_F(t))$. A simple calculation gives
\be
-2 \sinh \Bigl( {1\over 2} {\rd \omega_A\over \rd t}\Bigr) \CF^{(1)}_0(t) = e^{(1)}_0(t), 
\ee
and the right hand side can be read from (\ref{firstcorr}). We then find, 
\be
\CF^{(1)}_0(t)=-{\ri \over 4 \pi} {\rd e_F(t) \over \rd t} {1\over \sinh \Bigl( {1\over 2} {\rd \omega_A\over \rd t}\Bigr)}.
 \ee
 We can now use that 
 \be
 {\rd \omega_A\over \rd t}={\partial \omega_A \over \partial E}(e_F) {\rd e_F \over \rd t}=2\pi {{\partial \omega_A / \partial E} \over 
 {\partial \omega_B/ \partial E}} =2\pi\ri  { \oint_A {\rd x/p(x,e_F(t))} \over \oint_B {\rd x/p(x,e_F(t))}}
 \label{quotper}
 \ee
 to finally write 
 \be
  {\rd \omega_A\over \rd t} =2\pi \ri \tau (e_F(t)) ,
  \ee
  where $\tau(E)$, given by the quotient of periods in (\ref{quotper}), is the modulus of the curve defined by 
  \be
  \label{pcurve}
  y^2=p(x,E).
  \ee
 We then obtain an expression for $\CF^{(1)}_0(t)$ of the form,
  \be
 \CF^{(1)}_0(t)= -\Bigl( 2{\partial \omega_B\over \partial E}(e_F(t))   \sin \bigl( \pi \tau(e_F(t)) \bigr)\Bigr)^{-1},
 \ee
 which is written solely in terms of periods on the curve (\ref{pcurve}). In terms of the period of the trajectory $T$ and the tunneling time
 \be
 T_{\rm tun}=-{\partial \omega_A \over \partial E}
 \ee
 we can also write
 \be
  \CF^{(1)}_0(t)=\ri  \biggl[ 2 T(e_F)  \sinh \biggl( \pi {T_{\rm tun}(e_F) \over T(e_F)} \biggr)\biggr]^{-1}.
  \ee

One can derive an expression for $\mathcal{F}$ analogous to (\ref{Explicit}). First of all, let us note that the relation (\ref{exactrel}) can be written as
\begin{equation}
 \mathcal{F}=\frac{1}{2g_s\sinh\left(
\frac{g_s}{2}\,\frac{\d}{\d t}\right)}\,e(t,g_s)\;.
\end{equation}
Then
\begin{equation}
 \mathcal{F}=\frac{1}{2g_s D\sinh(\pi g_s D)}:e^{-D\omega^{\rm q}_B(g_s,e_F)}(1+a^A(g_s,e_F))^{\ri g_s D}:\,\frac{1}{T(e_F)}\;,
\end{equation}
where 
\be
D=\frac{1}{T(e_F)}\frac{\d}{\d e_F}.
\ee
The $\ell$-instanton contribution is
\begin{equation}
   \mathcal{F}^{(\ell)}=\frac{\ri}{2\,\ell!}\,\frac{P^{(\ell)}(g_s D)}{\sinh(\pi g_s D)}:e^{-D\omega^{\rm q}_B}:\frac{(a^A)^\ell}{T}\;,
\end{equation}
with one-loop part
\begin{equation}
  \mathcal{F}^{(\ell)}_0=\frac{\ri}{2\,T\,\ell!}\,\frac{P^{(\ell)}\left(\frac{T_\tun}{T}\right)}{\sinh\left(\pi\frac{T_\tun}{T}\right)}\;.
\end{equation} 

 \subsection{Large order behavior of the $1/N$ expansion}
 
 By standard arguments \cite{lgzj}, the large order behavior of the $1/N$ expansion of the ground state energy should be governed by the 
 instanton corrections that we have computed. Therefore, we can test our results by comparing them to the behavior of the 
 amplitudes $\CF_g(t)$, as $g$ becomes large. We will restrict ourselves to the case of the unstable potential. The double-well potential, which 
 is slightly subtler, can be studied in a similar way (see for example \cite{zjj}). 
 
 Let us write the one-instanton amplitude as
 \be
 \label{onelarge}
 g_s^2 \CF^{(1)}(t,g_s)= \ri g_s^{-b} \re^{-\frac{S}{{g_s}}} \sum_{n=0}^{\infty} \mu_n (t) g_s^{n}.
 \ee
Then, the perturbative amplitudes should have the large order behavior (see for example \cite{mswone})
\be
\CF^{(0)}_g \sim {S^{-2g-b} \over \pi}\, \Gamma(2g+b)\biggl( \mu_0 + \sum_{n\ge 2} {\mu_n S^{n-1} \over \prod_{k=1}^{n-1} (2g+b-k)} \biggr).
\ee
 In (\ref{onelarge}) the one-instanton amplitude is
the discontinuity across the positive real axis, i.e. the difference  
between the results obtained with the right and the left Borel
resummations. In our case, the left resummation is purely  
perturbative, and we can use the results for the one-instanton  
amplitude of section 3.2. We also have that $b=-2$, $S=\omega_A(e_F(t))$, and we obtain the leading asymptotics
\be
\label{leadas}
\CF^{(0)}_g (t) \sim { \Gamma(2g-2) (\omega_A(e_F))^{-2g+2} \over 2 \pi  T(e_F)  \sinh \bigl( \pi T_{\rm tun}(e_F) / T(e_F) \bigr)} 
\ee

\begin{figure}[!ht]
\leavevmode
\begin{center}
\includegraphics[height=4cm]{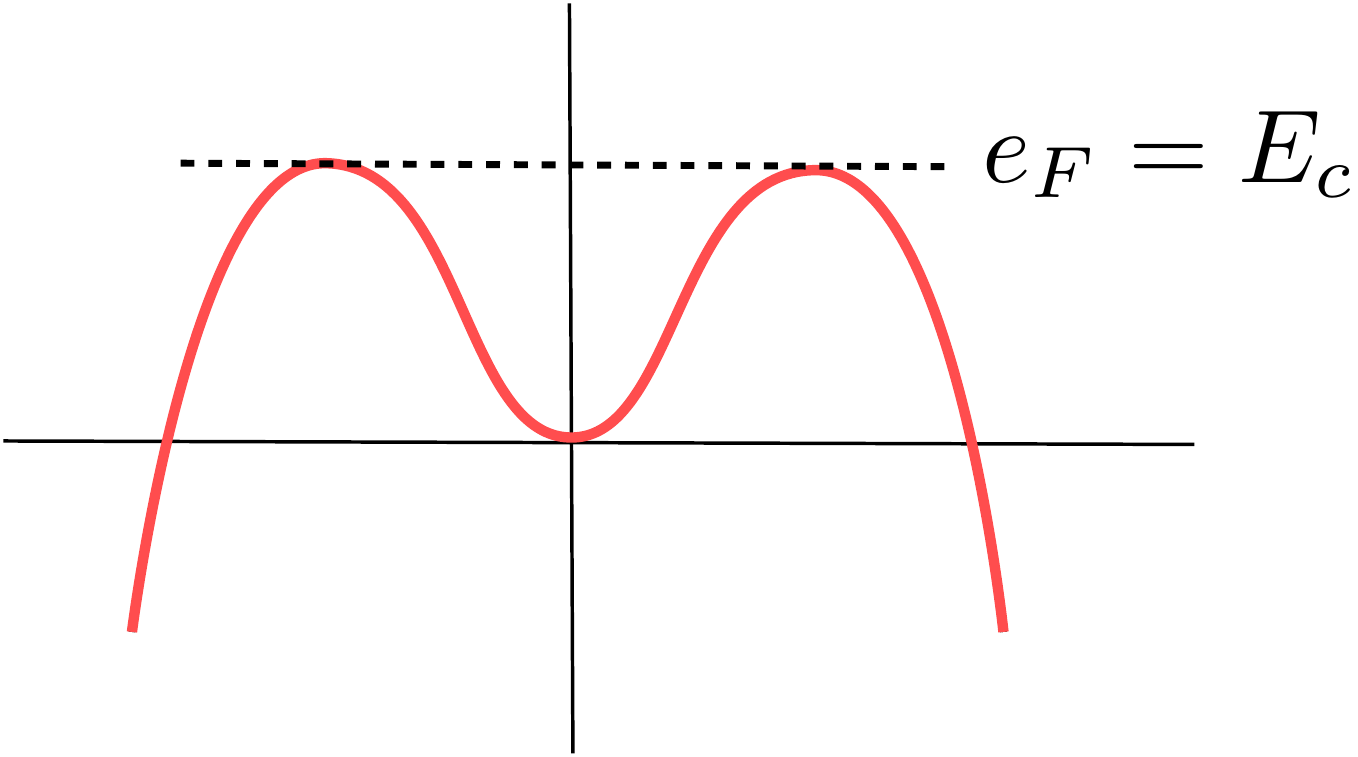}
\end{center}
\caption{The critical point of Matrix Quantum Mechanics in the inverted quartic potential takes place when the Fermi level $e_F$ reaches the maximum $E_c$.}
\label{critical}
\end{figure} 

We will now present test this formula with the inverted quartic potential (\ref{gquartic}) that we discussed above and its double-scaling limit. We will fix 
the normalization by choosing $g=-1$. The energy is then related to the modulus $k^2$ introduced in (\ref{inversemod}) by
\be
E(k)={1\over 4} {k^2 \over (1+k^2)^2},
\ee
therefore the relation $E(k)=e_F(t)$ gives an implicit relationship between $t$, the 't Hooft parameter, and the modulus $k$. When $k^2=1$ we reach a critical point and the $1/N$ expansion 
breaks down. Physically, this critical point occurs when the Fermi level $e_F(t)$ attains the maximum of the potential $E_c$, as shown in \figref{critical}. 
This critical point plays an important 
r\^ole in non-critical string theory, since it makes possible to define the $c=1$ string by a double-scaling limit, see \cite{gm,klebanov} for reviews.

\begin{figure}[!ht]
\leavevmode
\begin{center}
\includegraphics[scale=0.6]{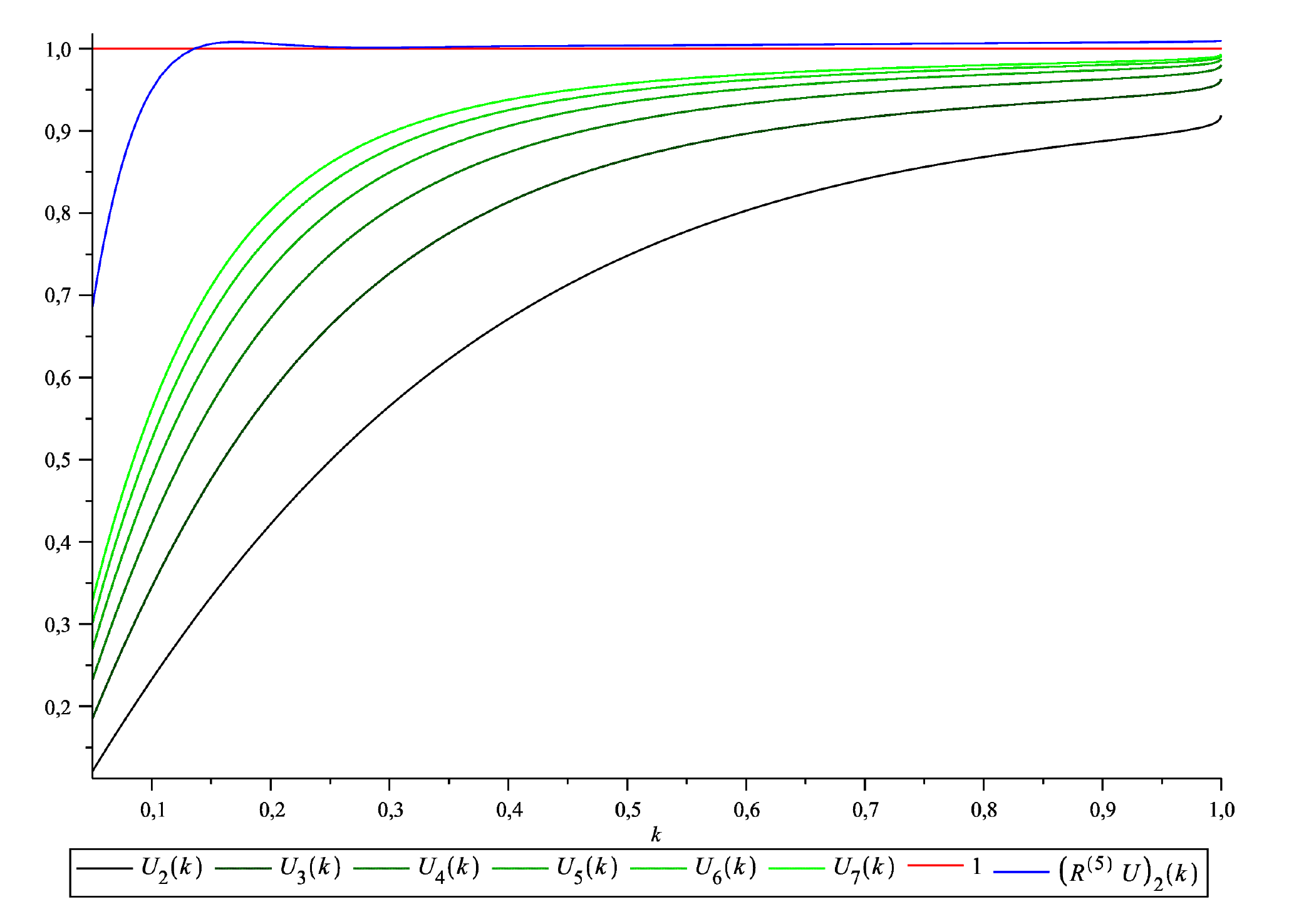}
\end{center}
\caption{In this figure we plot the first functions $U_g(k)$ and the second element of the fifth Richardson transform $(R^{(5)}U)_2(k).$}
\label{plot}
\end{figure}

We have computed $\CF^{(0)}_g(t)$ up to $g=7$ in order to test the large order behavior (\ref{leadas}). Since the potential is symmetric, we have 
two identical contributions from instantons going from $a$ to $b$, and from instantons going from $-a$ to $-b$, therefore in this case we have to add an extra factor of $2$ in (\ref{leadas}). To do the test, we notice that the sequence
\begin{equation}
 U_g(E)= \pi\, T(E)\,\sinh\left(\pi\,T_{\rm tun}(E)/T(E)\right) \,\frac{\left(\omega_{A}(E)\right)^{2g-2}}{(2g-3)!} \, \CF^{(0)}_g
\end{equation} 
should approach $1$ as $g\rightarrow \infty$, for all $0<\kappa<1$. Of course we have only a few points to test this, but we can use Richardson transforms to accelerate 
 the convergence. We recall that, given a sequence $U_g$, $g=0, \cdots, L$, the $r$-th Richardson transform gives a sequence of $L+1-r$ terms 
 \be
\left( R^{(r)} U\right)_n=\sum_{m\geq 0}{U_{n+m}(n+m)^r(-1)^{m+r}\over m!(r-m)!}  
\ee
with accelerated 
 convergence. In \figref{plot} we plot the first few functions $U_g(k)$ as well as $(R^{(5)}U)_2(k)$. The agreement is quite good, and confirms our analytic derivation of the 
 one-instanton correction. 
 
\subsection{Application to the $c=1$ string}

 As we mentioned above, if we consider the inverted double-well potential with $k\rightarrow 1$, we can define a double-scaled theory by considering the limit
\begin{equation}
 E_c-e_F=\mu\rightarrow 0\;\;,\;\;\ {g_s \over \mu}= z \quad \text{fixed},
\end{equation}
This limit defines the $c=1$ string, and the function $\CF^{(0)}(t,g_s)$ becomes (see, for example, \cite{klebanov})
\be
\label{dsenergy}
\CF^{\rm ds}(z)={1\over 8\pi} \biggl\{ -4  {\log z \over z^2} + {1\over 3} \log z +\sum_{g=2}^{\infty} \frac{(2^{2g-1}-1)|B_{2g}|}{(g-1)g(2g-1)2^{2g-2}} z^{2g-2} \biggr\}.
\ee
Therefore, in this limit the asymptotics as $g\rightarrow \infty$ can be computed directly by using that
\be
|B_{2g}|\sim\frac{2(2g)!}{(2\pi)^{2g}}, \qquad g\gg1. 
\ee
The genus $g$ coefficient of (\ref{dsenergy}), which we will denote by $\CF^{\rm ds}_g$, behaves like
\begin{equation}
\label{exas}
  \CF^{\rm ds}_g \sim \frac{1}{2\pi^3}\,(2\pi)^{-2g+2}\Gamma(2g-2)\;
\end{equation}
Let us compare this to the predictions of (\ref{leadas}). The different quantities involved in that expression can be easily computed in the 
double-scaling limit. Since\footnote{Here we actually use a different normalization of the potential (\ref{gquartic}) with $g=-\frac{1}{9}$.}
\be
E_c-V(\lambda)=-\frac{1}{2}(\lambda-\lambda_*)^2+\cdots
\ee
 Denoting $\xi=\lambda-\lambda_*$ one computes 
\begin{equation}
\ba
 T_{\rm tun}(e_F)&=4\int\limits_0^{\sqrt{2\mu}}\frac{\rd\xi}{\sqrt{2\mu-\xi^2}}=2\pi,\\
 {1\over g_s} \omega_{A}(e_F)&={4\over g_s} \int\limits_0^{\sqrt{2\mu}}{\rd\xi}{\sqrt{2\mu-\xi^2}}=2\pi{\mu \over g_s}=\frac{2\pi}{z}.
\ea
\end{equation}
Since $T(e_F)$ diverges in this limit, 
\be
T(e_F)  \sinh \biggl( \pi {T_{\rm tun}(e_F) \over T(e_F)} \biggr) \rightarrow \pi T_{\rm tun}(e_F). 
\ee
After including the extra factor of $2$ to account for the symmetry of the potential (which is also 
included in the formula of \cite{klebanov}), we find that (\ref{leadas}) reproduces (\ref{exas}). We should mention that 
already in the paper \cite{gzj}, the leading contribution to the large order behavior of the $c=1$ string is determined by computing the action of the instanton in the 
double-scaling limit. The paper \cite{ak} also discusses nonperturbative effects in the $c=1$ string.

\sectiono{Conclusions and open problems}

In this paper we have determined the non-perturbative corrections to the ground state energy in large $N$ Matrix Quantum Mechanics. Essentially, we reduced 
the problem to the calculation of multi-instanton corrections in conventional Quantum Mechanics, which can be obtained in turn through
exact quantization conditions. 

There are various possible generalizations of our work. We have restricted ourselves to the ground state energy, but the nonperturbative corrections to 
energies of excited singlet states or non-singlet states (like the adjoint state analyzed in \cite{mo}) could be also analyzed with our techniques. It would be also 
very interesting to analyze the large $N$ multi-instantons once 
fermionic degrees of freedom have been introduced, as in \cite{affleck}. Finally, we have 
worked out in detail the example of the unstable quartic potential and its double scaling limit, and we have 
given all the necessary ingredients to understand the double-well potential. It would be interesting to work out this 
example in more detail, since is relevant to the understanding of type 0B strings \cite{toumbas,hat}. 

\section*{Acknowledgements}
This work was supported in part by the Fonds National Suisse. 

\appendix 

\sectiono{Resurgence and nonperturbative quantization conditions} \label{resurgence}

In this Appendix we briefly review the nonperturbative treatment of the Schr\"odinger equation based on the theory of resurgence and described in detail in \cite{ddp,ddptwo}.

In this approach, the WKB expansions satisfying the Schr\"{o}dinger equation, such as (\ref{WKB_wave}), as well as other relevant functions written in terms of series in $\hbar$, are 
regarded as so-called resurgent symbols, i.e. formal sums of the form 
\begin{equation}
 \varphi=\sum_{\omega}\varphi_\omega e^{-\frac{\omega}{\hbar}}
\end{equation}
where $\varphi_\omega$ are formal series in $\hbar$:
\begin{equation}
 \varphi_\omega=\sum_{n=0}^{\infty}a_n^\omega\hbar^n\;.
\end{equation}
 The formal power series $\varphi_\omega(\hbar)$ must satisfy the condition that their Borel transforms
\begin{equation}
 \widehat{\varphi}_\omega(\xi)=\sum_{n=1}^\infty\,a_n\,\frac{\xi^{n-1}}{(n-1)!}
\end{equation} 
have only finite number of singularities in the positive real direction, so that the inverse transform (which is called a resummation of the series 
$\varphi_\omega(\hbar)$) can be defined by
\begin{equation}
 a_0+ \int\limits_{\gamma}e^{-\frac{\xi}{\hbar}}\hat{\varphi}_\omega(\xi)d\xi     \label{inverse_Borel}
\end{equation} 
where $\gamma$ is some contour starting from zero and going along the positive real direction which avoids these singularities\footnote{If nothing is known about the growth of $\widehat{\varphi}_\omega(\xi)$ at infinity, the function (\ref{inverse_Borel}) can be defined only up to a function $f(\hbar)$ of hyperexponential decrease, i.e. $\forall k,\;\exists C\;:\;|f|<C\,e^{-\frac{k}{\hbar}}$.}. There are two natural choices of $\gamma$: the contour $\gamma_-$ avoiding the singularities from the left, and the contour $\gamma_+$ avoiding the singularities from the right. The corresponding results give us the left and right resummations: $s_-\varphi_\omega(\hbar)$ and $s_+\varphi_\omega(\hbar)$. One can construct from them the left and the right resummations of the resurgent symbol $\varphi$:
\begin{equation}
 s_-\varphi=\sum_{\omega}s_-\varphi_\omega e^{-\frac{\omega}{\hbar}}\;,\;s_+\varphi=\sum_{\omega}s_+\varphi_\omega e^{-\frac{\omega}{\hbar}}
\end{equation} 
These resummation operators $s_+$ and $s_-$ actually define \textit{isomorphisms} 
of the algebra of resurgent symbols into the algebra of so-called extended resurgent functions. 
The linear space of WKB symbols will be denoted by $\WKB$. The two-dimensional linear space of solutions to the Schr\"{o}dinger equation, regarded as extended resurgent functions, will 
be denoted by $\mathcal{S}$. Notice that the same resurgent function can be obtained from two different symbols by using the different resummations $s_\pm$. 
One can define the action of the Stokes automorphism $\mathcal{S}$ on $\WKB$ by requiring the commutativity of the following diagram:
\begin{equation}
\xymatrix{
\WKB \ar[rr]^\Stokes \ar[rd]_{s_-} & &\WKB \ar[ld]^{s_+} \\
& \mathcal{S} 
}\;\;.
\end{equation}

It turns out that WKB-symbols are well defined only in the so-called Stokes regions of the complex $x$-plane. We will denote the space of WKB symbols in the Stokes region $R$ by $\WKB(R)$. The complex plane is divided into Stokes regions by Stokes lines. Stokes lines are lines starting or ending (or both) at critical points (zeroes of the momentum $p(x,E)$) along which $\re^{\frac{i}{\hbar}\int^xp(\xi,E)d\xi}$ decreases/increases fastest.
Inside a Stokes region the space of WKB-symbols can be decomposed into the direct sum with respect to the choice of sign of the momentum ($\pm p(x,E)$): 
\be
\WKB(R)=\WKB^{(+p)}(R)\oplus \WKB^{(-p)}(R).
\ee
The Stokes automorphism respects this decomposition inside the Stokes region $R$. 

Consider now two different Stokes regions $R$ and $R'$ and choose some resummation prescription ($s_+$ or $s_-$). Then there is a map 
\be
\mathcal{C}^\pm_{RR'}\,:\,\WKB(R)\rightarrow\WKB(R')
\ee
 called the connection isomorphism, which relates two different WKB-symbols corresponding to the same global function through the different resummation $s_\pm$ . 

Let us now consider a simple pattern of Stokes lines: three unbounded Stokes lines $L,\,L',\,L''$ going out from one critical point:
\begin{equation}
 \xymatrix{
L & R' &  \\
  R \ar@/^/[ur]^(0.7){\mathbf{L}} \ar@/_4pc/[ur]_{\delta_L} 
& *=0{\bullet}\ar@{-}[ul]\hole \ar@{-}[r] \ar@{-}[dl] 
& L'  \\
 L'' & & 
} \label{elementary_connection}
\end{equation} 
One can decompose
\be\WKB(R)=\WKB^{(+p)}(R)\oplus \WKB^{(-p)}(R)\ee
and 
\be\WKB(R')=\WKB^{(+p)}(R')\oplus \WKB^{(-p)}(R'),\ee
where $\WKB^{(+p)}$ consists of symbols dominant 
along $L$, i.e. with the factor 
\be
\exp\left( \frac{\ri}{\hbar}\int^x p(\xi,E)\rd\xi\right)
\ee
 increasing along $L$, and $\WKB^{(-p)}$ consists of symbols recessive along $L$, 
 i.e. with the factor 
 \be
 \exp\left(-\frac{\ri}{\hbar}\int^xp(\xi,E)\rd\xi\right)
 \ee
 decreasing along $L$. 
Then 
the elementary 
connection 
isomorphism can be written 
in 
the 
following matrix 
form:
\begin{equation}
 \mathcal{C}^\pm_{RR'}=\left(\begin{array}{cc}
\mathbf{L} & 0\\
\delta_L & \mathbf{L}                        
                       \end{array}\right)\;,
\end{equation}
where $\mathbf{L}$ is analytic continuation across $L$ and $\delta_L$ is analytic continuation along the contour around $L$, as shown in (\ref{elementary_connection}). The elementary connections operators for the right and left resummations are the same. 

\begin{figure}[!ht]
\leavevmode
\begin{center}
\includegraphics[scale=0.7]{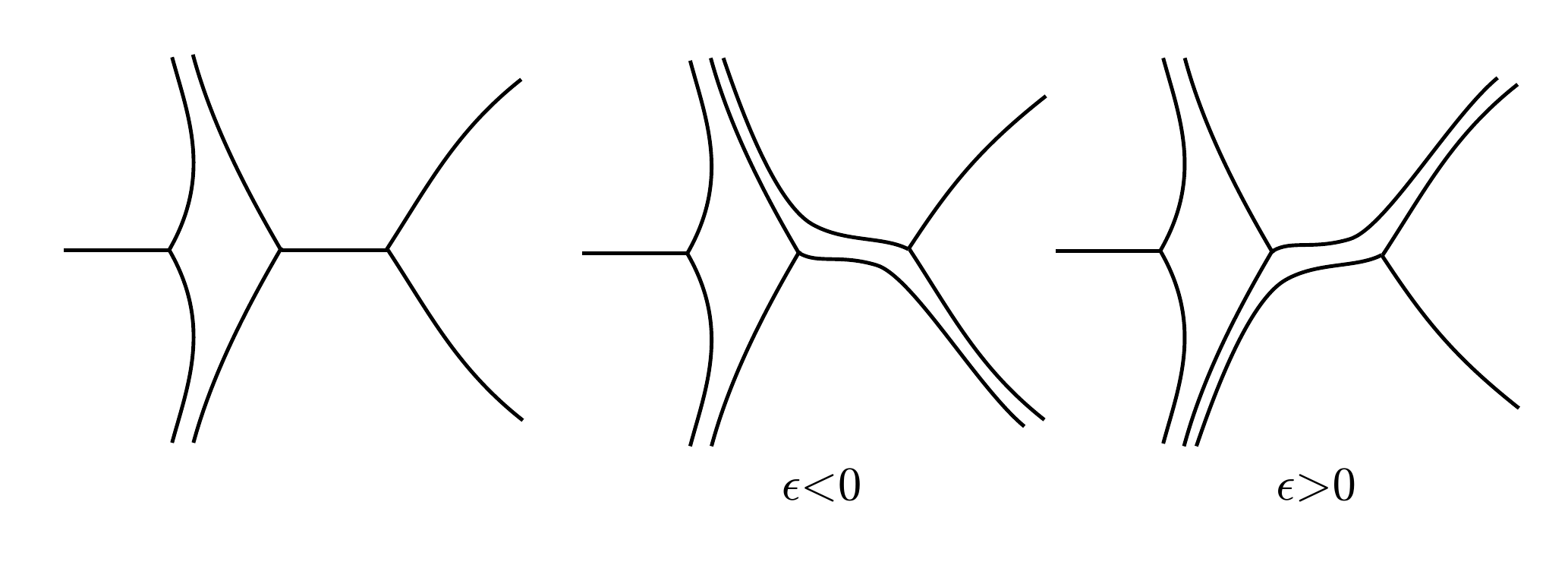}
\end{center}
\caption{Stokes lines for the cubic potential and their resolution.}
\label{stokes_cubic}
\end{figure}

The global picture of Stokes lines can not be always represented as a composition of simple patterns as the one depicted in (\ref{elementary_connection}). In general one has bounded Stokes lines. However, one can always make small deformations such that all Stokes lines will be unbounded, and one can then find connection isomorphisms $C_{RR'}^\pm$ between every two Stokes regions $R,\,R'$ by using the elementary connection isomorphisms that we just described. Deformed Stokes lines are lines along which 
\be
\exp\left( \frac{\ri \re^{-\ri\epsilon}}{\hbar}\int^xp(\xi,E) \rd\xi\right)
\ee
decreases/increases fastest, where $\epsilon\rightarrow 0$, see \figref{stokes_cubic}. One can use the deformation with $\epsilon>0$ (respectively, $\epsilon<0$) to compute the connection isomorphism for the right (resp., left) resummation.

The problem of finding energy levels can be formulated as finding the values of the energy $E$ such that the subspace $\mathcal{S}_{-\infty}\subset\mathcal{S}$ of solutions descending at $+\infty$ (or having negative momentum if one searches for resonances in unstable potentials) coincides with the subspace of solutions $\mathcal{S}_{+\infty}\subset\mathcal{S}$ descending at $+\infty$ (having negative momentum in the case of resonances). This is equivalent to the vanishing of the so-called Jost operator $\mathcal{J}\,:\; \mathcal{S}_{-\infty}\rightarrow \mathcal{S}/\mathcal{S}_{+\infty}$. If one chooses the basis (Jost basis) $(\phi,\phi^\ast)$ of $\mathcal{S}$ such that $\phi\in\mathcal{S}_{-\infty}$ and $\phi^\ast\notin\mathcal{S}_{+\infty}$, the action of the Jost operator is given by a Jost function $J(E)$ defined by
\begin{equation}
 \phi=J(E)\phi^\ast\;\mod\;\mathcal{S}_{+\infty}\;.
\end{equation}
Then the energy levels are given by the equation $J(E)=0$. 

Let us consider the WKB-symbol $\varphi$ defined in the Stokes region $R_1$ unbounded in the negative real direction, such that $\phi$ is the resummation of $\varphi$. If $\phi^\ast$ is the resummation of the WKB-symbol $\tilde{\varphi}$ defined in the Stokes region $R_2$ unbounded in the positive real direction, we can write
\begin{equation}
\mathcal{C}_{R_1R_2}^\pm\varphi=J^{\pm}(E)\tilde{\varphi} \quad \mod s_\pm^{-1}\mathcal{S}_{+\infty}. \end{equation} 
$J^{\pm}(E)$ is called the right/left resurgent symbol of the Jost function. Correspondingly, the right/left symbols of the energy levels are the roots of the equation $J^\pm(E)=0$. It is usually convenient to represent Jost symbols in the form 
\be
J^\pm=1+\re^{\frac{\ri}{\hbar}\Omega^{\pm}(E)}.
\ee
The nonperturbative quantization condition is then
\begin{equation}
 \Omega^\pm(E)=2\pi\chi\;. \label{nopert_Omega_pm}
\end{equation} 
Thus the problem of finding a nonperturbative equation for the energy reduces to finding the connection isomorphism $\mathcal{C}^\pm_{R_1R_2}$ using the rule (\ref{elementary_connection}). In many examples, the physically relevant resummation prescription is the median resummation defined by 
\be
s_\med=s_-\circ\Stokes^{-\frac{1}{2}}=s_+\circ\Stokes^{\frac{1}{2}}.
\ee
 The corresponding median Jost symbol is 
 \be
 J^{\med}=\Stokes^{\frac{1}{2}}J^-=\Stokes^{-\frac{1}{2}}J^+.
 \ee
 This prescription has the property that, in the case of a stable potential, the solutions to the equation 
 \be
 J^\med(E)=1+\re^{\frac{\ri}{\hbar}\Omega_\med}=0
 \ee
  are resurgent symbols with real coefficients.

\end{document}